\documentclass[12pt,a4paper]{article}

 \usepackage[dvips]{graphics}
 \usepackage{graphicx}
 \usepackage{afterpage}
 \usepackage{enumerate}
\begin{document}

\title{\bf Convective shifts of iron lines in the spectrum
 of the solar photosphere}
 \author{\bf V.A. Sheminova and A. S. Gadun}
 \date{}

 \maketitle
 \thanks{}
\begin{center}
{Main Astronomical Observatory, National Academy of Sciences of
Ukraine
\\ Zabolotnoho 27, 03689 Kyiv, Ukraine\\ E-mail: shem@mao.kiev.ua}
\end{center}

 \begin{abstract}
The influence of the convective structure of the solar
photosphere on the shifts of spectral lines of iron was studied.
Line profiles in the visible and infrared spectrum were
synthesized with the use of 2-D time-dependent hydrodynamic
solar model atmospheres. The dependence of line shifts on
excitation potential, wavelength, and line strength was
analyzed, along with the depression contribution functions. The
line shifts were found to depend on the location of the line
formation region in convective cells and the difference between
the line depression contributions from  granules and
intergranular lanes. In visible spectrum the weak and moderate
lines are formed deep in the photosphere. Their effective line
formation region is located in the central parts of granules,
which  make the major contribution to the absorption of
spatially unresolved lines. The cores of strong lines are formed
in upper photospheric layers where is  formed  reversed
granulation due to convection reversal and physical conditions
change drastically there. As a consequence the depression
contributions in the strong line from intergranular lanes with
downflows substantially increase. This accounts for smaller blue
shifts of strong lines. In infrared spectrum the observed
decrease in the blue line shifts is explained by the fact that
their effective line formation regions lie higher in the
photosphere and extend much further into the reversed
granulation region due to the opacity rise with the increase of
line wavelength.  Additionally the effective line formation
depths of the synthesized visible and infrared Fe~I lines and
their dependence on line parameters is discussed.
\end{abstract}

\section{Introduction}

If the motion of the Sun-Earth system is taken into account as
well as  the correction for the gravitational red shift (636
m/s) caused by the difference between the gravitational
potential of the solar atmosphere and the Earth is introduced,
the  most absorption lines observed at the center of the solar
disk are shifted  towards shorter wavelengths  with respect to
laboratory lines. The observed blueshifts of the solar iron
lines are 300--400 m/s on the average \cite{9,1,11,12,13,26,28,32}.
They were measured to an accuracy of 100 m/s in the range from 0
to 1 km/s. It is found that the blueshifts decrease with growing
height of line formation as suggested by its dependence on
excitation potential.  The largest blueshifts are observed in
weak lines with high excitation potentials, while the smallest
blueshifts are in strong lines. The redshifts were found to be
small in some strong lines in the visible spectrum, but they can
be as large as 1 km/s in the ultraviolet lines in the range
195--200 nm \cite{32}. Some ``reddening'' of blueshifts was also
noted in active photospheric regions  \cite{1,11}. The blueshifts
observed in weak and moderate lines can be explained by a
decrease with height in the velocity of the convective motions
along the line of sight towards the observer  \cite{12,13}. The
formation region of strong photospheric lines extends to the
layers of  reversed granulation, a pattern comparable  to the
granulation but with reversed brightness modulation. The matter
at the center of convection cells becomes cooler than in
descending intergranular lanes, but it still moves upwards. An
inverse correlation between brightness and shifts was found in
the cores of very strong lines in the visible and infrared
spectrum  \cite{22}. The effects of convection overshoot in the solar
photosphere  \cite{27} is believed to be the cause of the observed
line redshifts.

The objective of this study is the effect of reversed
granulation and the role of granules and intergranular lanes in
the formation of spatially unresolved spectral lines and the
cause of the ``reddening'' of blueshifts of strong lines. Such a
study is possible today due to the remarkable progress in the
simulation  of solar granulation and its evolution
\cite{10,14,27,30,33} and in the development of the idea of
depression contribution functions  \cite{19,20,23,25}. Based on
2-D hydrodynamic model atmospheres  \cite{18}, we synthesized
Fe~I lines and calculated their shifts. We used contribution
functions to investigate the region of line formation in
convection cells and to analyze the influence of physical
conditions in the granular photosphere on the shifts of the
synthesized lines.

\section{Hydrodynamic model of solar granulation}

The sequence of single-scale (one-cell) and multi-scale
two-dimensional nonstationary hydrodynamic  (HD) models of solar
granulation built by Gadun  \cite{2,18} were used more than once
for the synthesis of photospheric lines
\cite{3,4,6,7,8,16,5,15,17}. Such sequence models proved to be
capable of reproducing satisfactorily the manifestations of
convective motions, the asymmetry and shifts of lines in
particular. Comparison of the profiles synthesized with the use
of 3-D and 2-D HD models  \cite{2} showed that the single-scale
2-D models are inferior to 3-D models in reproducing the
observed line asymmetry because their inhomogeneity spectrum is
narrower. Nevertheless, the use of 2-D models  \cite{18} for the
synthesis of spectral lines is justified when we want to obtain
the overall picture of the processes in the superadiabatic
regions in the photospheres of the Sun and solar-like stars.

The 2-D HD model sequence we used in our study treats the
radiative transfer on the basis of the monochromatic absorption
coefficient depending on frequency, and the atomic and molecular
line absorption is taken into account through the use of the
opacity distribution function (ODF) method  \cite{24}. The
thermal convection in the solar envelope was assumed to be
quasi-stationary with a single convective flow in the simulation
region, and this flow represented the only granulation scale. In
building the HD models we assumed that convective flows were
quasi-stationary in time, the horizontal size of convective
cells did not change in the course of their evolution, and
convective cylinders always extended downwards as deep as
permitted by the simulation region (1100 km below the surface
level in the models). These models were called the steady-stable
(SS) models by their author. The dimensions of the calculated
region were 1400 km in the horizontal direction and 1960 km in
the vertical direction (the atmospheric layers extended to 800
km). The spatial step was 28 km. The total duration of
simulation was 50 min of the real solar time with a time step of
30 s between the models. Test calculations showed that such a
duration was sufficient for a correct representation of
five-minute oscillations in the spectral line calculations. This
time-dependent sequence of 2-D models for the solar disk center
includes 100 models (or 100 snapshots) consisting of 48 columns.
Each column is located along the line of sight. A more detailed
description of 2-D HD models of this type can be found in
\cite{3,18}.

\section{Spectral line synthesis}

To find the dependence of line shifts on principal line
parameters, we used certain artificial Fe~I lines. The central
line depths ($d$), lower excitation potentials ($EP$), and
wavelengths of the artificial lines was specified in a wide
range of actual Fe I line parameters. Then the product of
abundance and oscillator strength $Agf$ was chosen for every
line in accordance with its central depth.  The lines were
synthesized in the 1.5-D LTE solution of the transfer equation
\cite{spansat} with the use of the SS-model sequence -- this
means that the line profiles were calculated for each column in
all snapshots as for a plane-parallel atmosphere. Then the
obtained line profiles were  averaged over all columns in each
snapshot and over all snapshots. The result was a line profile
averaged over space and time.

The central wavelength of a synthetic profile $\lambda$ was
determined by fitting a fourth-degree polynomial to the profile
core. The accuracy of the procedure was $\pm50$ m/s. The shift
of the synthetic line with respect to the laboratory central
wavelength $\lambda_0$ was calculated, with allowance made for
the gravitational shift, in radial velocity units by the formula
$ V_R =c( \lambda-\lambda_0 ) / \lambda_0-2.12 \cdot 10^{-6} c$,
$c$ being the light velocity. Test calculations demonstrated
that the shift of the averaged line profile synthesized with the
full SS-model sequence (100 snapshots) differs insignificantly
from the shift of the line profile calculated with a snapshot
averaged over all 100 snapshots. This allowed us to
substantially cut the calculations by using the averaged SS
model only. For line shift analysis we used the calculated line
profiles for each column in the averaged SS model and the
profile averaged over all columns. We also used the depression
contribution functions calculated for each profile point and the
effective formation heights. These latter are calculated as the
weighted mean for the depression contribution function
\cite{19,20}.

\section{Analysis of calculation result for lines in the visible
spectrum}

The plots of line shifts of synthetic Fe I lines for various
wavelengths, line strengths, and heights of formation (Fig. 1)
suggest that these calculated shifts behave in the same fashion
that the observed shifts  \cite{9,1,13}. Weak lines with high $EP$  at
shorter wavelengths have the largest blueshifts that correspond
to the largest negative radial velocities. Strong lines, whose
central parts are formed high in the photosphere, feature
redshifts. The typical strong narrow lines with $d=0.8$,
equivalent width  $W=8$~pm, and $\lambda=600$~nm show the
largest redshifts (positive radial velocities). Our results also
demonstrate a clear dependence of shifts on wavelength (Fig.
1a), although no such dependence was found in the observed line
shifts  \cite{9}. It seems likely that the authors could not reveal
this dependence because they ignored the differences in the
central depths of the lines observed.  In general, we confirm
that the line blueshifts decrease with growing line wavelength,
line strength, and effective height of line formation.

  \begin{figure}
\centerline{
\includegraphics [scale=0.3]{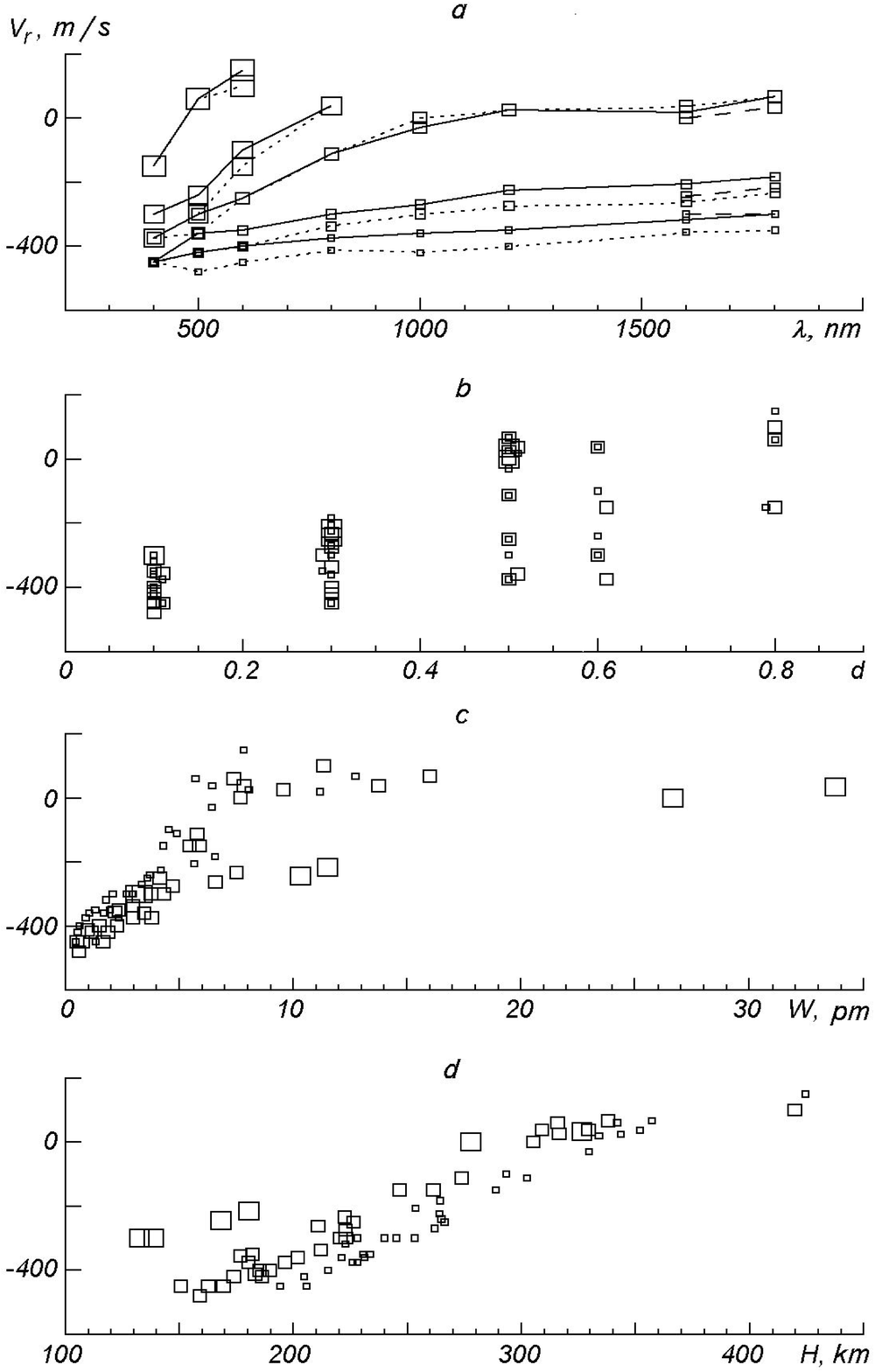}}
 \hfill
\vspace{0.1cm}
 \caption
{Dependence of convective shifts of synthesized Fe I lines on
wavelength (a), central depth (b), equivalent width (c), and
height of formation (d). Solid line: $EP=2.5$~eV, dotted line:
$EP=4.25$~eV, dashed line: $EP=6.4$~eV. Size of the squares
grows with line strength (a), $d=0.1$, 0.3, 0.5, 0.6, 0.8, and
excitation potential (b,c,d), $EP= 2.5$, 4.25, 6.4~eV.}
\label{F-1}

 \end{figure}

To understand the causes of formation line shifts, we considered
the snapshot of the simulated convective cell (the averaged SS
2-D model) and the depression contribution functions ($CF$) of
synthetic Fe I lines with different line parameters on Fig. 2.
The isotherms (dashed lines), which demonstrate the temperature
distribution in the convective cell, show that the temperature
inversion in the photospheric layers begins above 200 km. This
is the region of the so-called reversed granulation caused by
convective reversal. Here the isotherms, which are convex over
the granule center, become concave. The velocity field marked by
arrows illustrates the motions in the convective cell: vertical
upflows in the central part and downflows at the cell periphery.
The central region of convective cell  will be call granule,
and the regions with downflows will be call intergranular lanes,
as adopted for the observed granulation structure in the
photosphere. In addition, we also distinguish a region with
predominantly horizontal flows of matter located between the
granule and intergranular lanes. The depression contribution
function contours calculated in each column for the core of
synthesized line were plotted on the snapshot of the convective
cell from the averaged SS 2-D model. The solid thick line is the
contour corresponding $ 0.5 CF_{max}$. It outlines the region of
effective depression contributions. Inside this contour, there
are the contour of the greatest contributions corresponding
$0.75 CF_{max}$, and outside it two contours corresponding
$0.25CF_{max}$ and $0.1 CF_{max}$ are plotted. Here $CF_{max}$
is the maximum of contribution functions calculated in each
model column for line core. Outside the outermost contour the
line absorption is insignificant. Fig 2 shows  the contours for
the weak line (a) and strong line (b); for moderate line with
$EP =2.5$ (c) and $EP=4.25$ (d). Figs~2e,f,g,h show  the $CF$
contours for  moderately strong line depending on the
wavelength. It is interesting that lines with wavelengths about
$\lambda$~800~nm are only partially formed in the granulation
inversion zone, while the lines with $\lambda
\lambda$~1600--1800~nm are almost completely formed there. The
depression contribution from intergranular lanes is greater for
infrared lines. Figure~2h illustrates the appearance of new
small effective contribution regions in intergranular lanes.
  \begin{figure}
\centerline{
\includegraphics [scale=0.4]{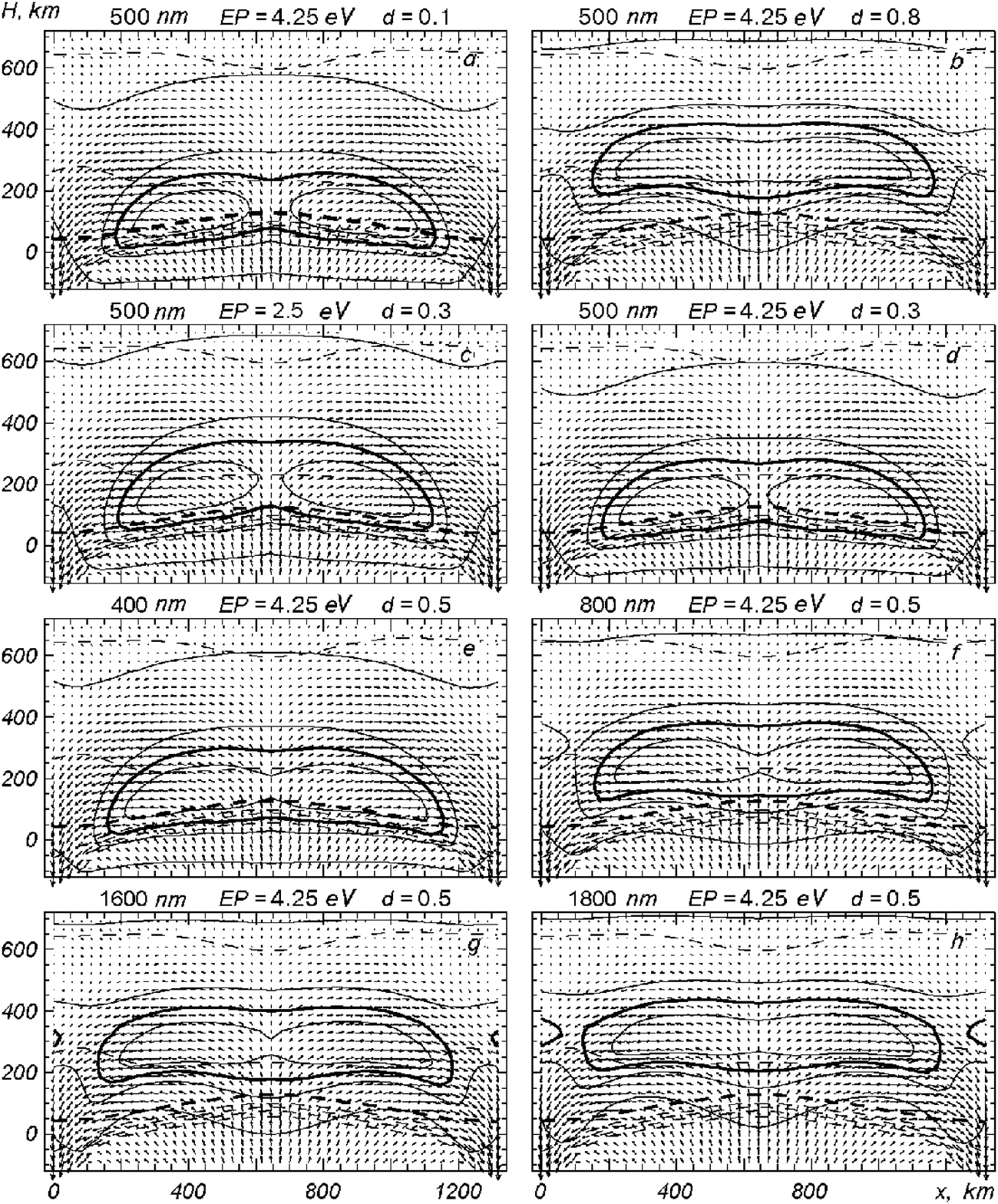}}
 \hfill
\vspace{0.1cm}
 \caption
{The snapshot of the convective cell from the averaged SS 2-D
model and contours of depression contribution functions (solid
line) for various synthesized Fe I lines: weak line and strong
line (a,b); moderate line with low and high excitation potential
(c,d);  moderate line with different wavelengths (e,f,g,h).
Thick solid line marks contour corresponding 0.5$CF_{max}$.
Dashed line: isotherms (4000, 4000, 6000, 7000, 8000, 9000,
10000 K from the top down), solid dashed line: surface level
with the Rosseland optical depth equal to unity. Arrows indicate
the velocity direction and their length is proportional to the
velocity amplitude.} \label{F-2}
 \end{figure}
In general, the depression contribution contours for different lines
plotted on the background of a snapshot of a convection cell
clearly demonstrated that intergranular lanes can play different
part in the formation of absorption lines depending on line
wavelength and  central line depth.

Apart from contribution functions, we also considered the
individual line profiles calculated for each column in the
simulated convective cell. We selected the most typical line
profiles formed in intergranular lanes, in region with
predominantly horizontal motions, and in the central parts of
the granule. Figure 3 displays these line profiles as well as
the profiles of the line formation  heights. It should be noted
that the line formation height profiles are quite intricate for
inhomogeneous photosphere models as different from the
homogeneous models. In the latter case these height profiles are
always completely symmetric. The synthetic line profiles at the
center of granule are slightly asymmetric and have always
blueshifts (kind I). The profiles in the regions with
predominantly horizontal motions have weak asymmetry and are
virtually not shifted (kind II). They differ insignificantly
from kind I profiles in strength and width. The profiles in
intergranular lanes are strongly asymmetric and have large
redshifts (kind III). Their cores are weaker and they are
markedly widened as compared to the profiles of kinds I and II.
Depending on the line parameters the shape and shift of the
profile of  I, II, III kinds slightly change, but its major
features remain the same. It should be noted that the line
profiles observed in the spectra obtained with very high spatial
and spectral resolution for bright, moderately bright, and dark
regions of photospheric granulation  \cite{21} are in complete
agreement with our synthetic profiles. This confirms to a degree
the correctness of our calculations with the use of 2-D HD
models.

Comparison of the profile averaged over the whole simulation
region   with the profiles of kinds I, II, and III (Fig. 3a)
shows that the largest contribution to the averaged profile core
of a weak line ($d=0.1$) comes from kind I and II profiles,
i.e., from the whole granule and  region with predominantly
horizontal motions. The contribution of the kind III profiles to
the average profile of the weak line is noticeable in the red
wing of the line only --- the wing is more extended, increasing
the profile asymmetry in extended wings. The blue wing of the
average profile of the weak line, as the line core, is formed in
the main by kind I and II profiles. It is therefore concluded
that the core shift and the shape of the blue wing of the weak
line are controlled primarily by the ascending motions in the
central parts of granules, while the shape of the red wing is
also affected by the descending motions in intergranular lanes.
From  Fig. 3c we can estimate the effective heights of formation
of the core and the wings in the averaged profile of a weak line
($d=0.1$). This is the layers in the range of 140--160 km for
the core, 80--140 km the blue wing, and 50--140 km for the red
wing.  The effective heights of the moderate line ($d = 0.5$) is
located 150 km higher then the above weak line. All other
details relative to different kinds of profiles are similar. In
the strong line ($d=0.8$, Fig. 3b) the kind III profiles formed
in intergranular lanes are much stronger, and the line asymmetry
is stronger because of temperature drop in the upper granular
layers. As a result, the shape of the unresolved profile of the
strong line change, and the blueshift of the line core
decreases. The  strong line core  is formed at heights of
280--350 km (Fig. 3d) in the reversed granulation region.
  \begin{figure}
\centerline{
\includegraphics [scale=0.3]{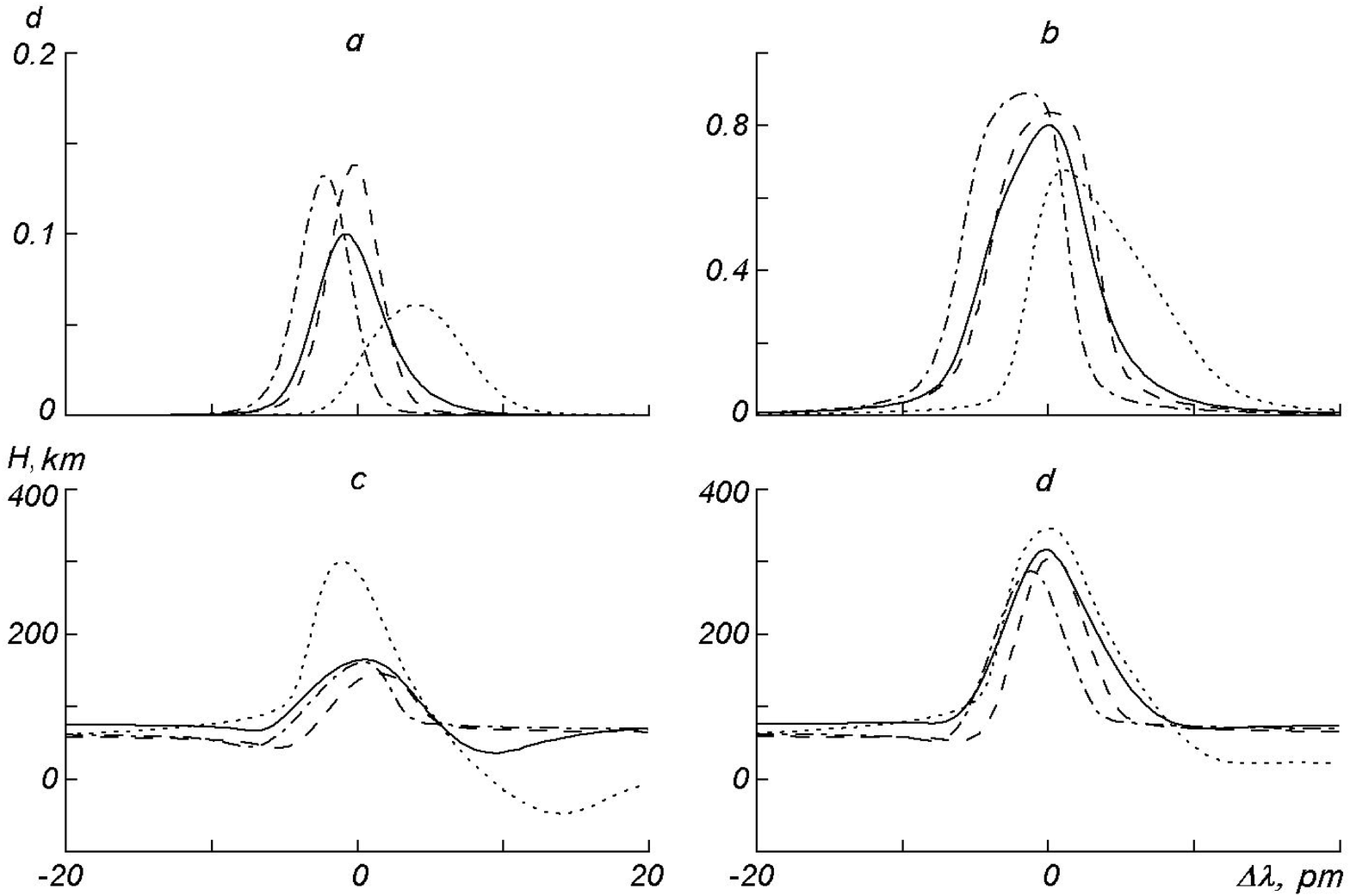}}
 \hfill
\vspace{0.1cm}
 \caption
{Three kinds of line profiles  calculated in  the simulated
convection cell for a weak line (a) and a strong line (b) at
$\lambda=500$~ nm, $EP=4.25$~eV. Dot-and-dash line: kind I
profile from the central region, dashed line: kind II profile
from the region with horizontal motions, dotted line: kind III
profile from intergranular lanes; solid line: average profile.
The profile of effective heights of line formation  (c,d) are in
response to changing line profiles (a,b).} \label{F-3}
 \end{figure}

The above analysis of the shape and shifts of synthesized line
profiles allows us to conclude that the red wing of spatially
unresolved line may be regarded as an indicator of the role
which intergranular lanes play in the formation of absorption
lines. The more extended red wind as compared to the blue wing,
the larger is the depression contribution in the spectral line
from interganular lanes. The shape of  blue wings and shifts of
weak and moderate lines in particular, depend completely on the
physical conditions in the central parts of granules.

\section{Analysis of infrared lines}

Lists of numerous unblended infrared lines in three bands of the
solar spectrum -- $J$ (1.00--1.34 $\mu m$), $H$ (1.5--1.8 $\mu
m$), and $K$ (1.9--2.5 $\mu m$) -- were published recently in
\cite{29,31}. The H band lines are most frequently used in various
spectral investigations of stellar atmospheres. These lines have
some advantages over other infrared lines. The continuum opacity
is at a minimum in the $H$ band, and we can see deeper
photospheric layers in these lines as compared to the lines of
$J$ and $K$ bands or the visible lines. For instance, the Fe I
lines with very high excitation potentials and large Lande
factors are very sensitive to the magnetic fields located at the
photosphere base. It is also well known that the contrast of
continuum intensity is smaller for infrared lines, and this
allows us to suggest that they are not formed in granulation
cells in the same way as the lines in the visible range. Some
differences between the profiles of the infrared and visible
lines with the same principal parameters are attributed to the
strong wavelength dependence of absorption coefficient. The
  \begin{figure}
\centerline{
\includegraphics [scale=0.30]{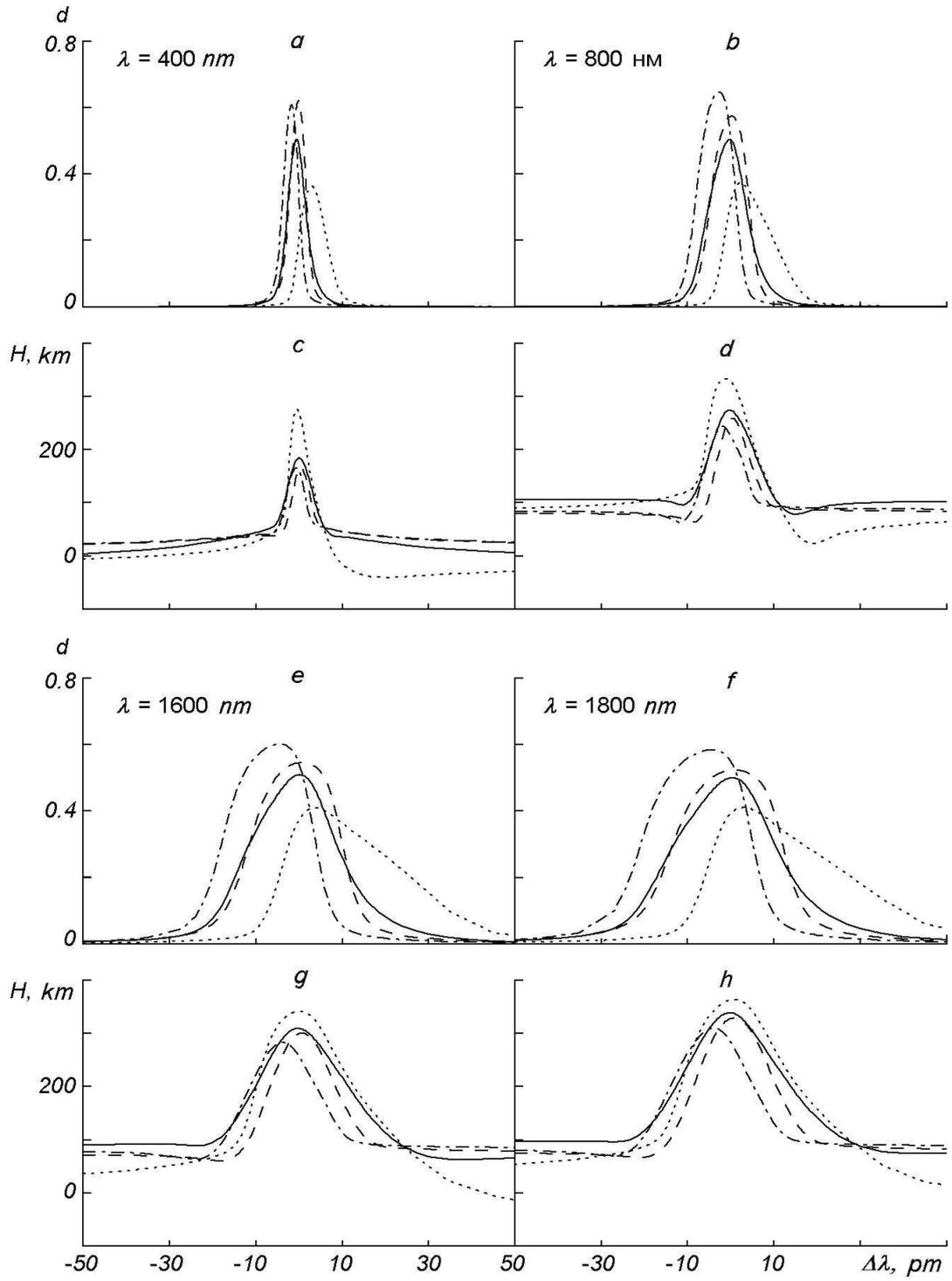}}
 \hfill
\vspace{0.1cm}
 \caption
{The same as in Fig. 3, but for moderate lines with $d=0.5$,
$EP=4.25$~eV, and $\lambda 400$ (a), $\lambda$800 (b), 
 $\lambda$1600 (e),  $\lambda$1800~nm
(f). } \label{F-4}
 \end{figure}
selective absorption grows proportionally to wavelength, while
the continuous absorption depends to a great extent on the sort
of absorbers. Recall that the continuum opacity in the infrared
range attains its maximum at $\lambda$~1000~nm, diminishes with
increasing wavelength to a minimum at $\lambda$~1600~nm which is
approximately of the same magnitude as the minimum at
$\lambda$~400~nm, and then grows again. Generally, the total
opacity is always greater in the infrared lines than in the
lines of the visible spectrum due to the increase of the
selective absorption coefficient. As a result, the regions of
formation of any infrared lines are located in higher layers and
are more extended into intergranular lanes. These effects
increase with line strength and excitation potential. The
excitation potentials for the infrared Fe I lines range from 2
to 6.4 eV, and for a line with $d=0.5$, $\lambda=1800$~nm, and
$EP=4.25$ eV, for instance, the region of its core formation is
located completely in the reversed granulation layers. In
intergranular lanes the line formation region shifts higher
(Fig. 2h) with respect to the visible lines of the same line
strength (Fig. 2e). The depression contribution from
intergranular lanes in the core of a visible line is about 25
percent of the maximum contribution, while for infrared lines it
is 50 percent and more.

The depression contributions of intergranular lanes to the
infrared line  affects the shape of average line profiles.
Figure 4 shows line profiles with a central line depth of 0.5
and wavelengths of 400, 800, 1600, and 1800 nm as well as their
formation height profiles. All infrared line profiles are nearly
twice as wide as the profiles of similar visible lines. The
widest and the most asymmetric profile is formed in
intergranular lanes, where the temperature and velocity
gradients are large, and as a result the asymmetry in the red
wings of average profiles is greater than in the visible lines.
The effective height profiles also suggest that the formation
regions of infrared lines are more extended in height. Weak
lines are formed in the range from $-100$~km to 300 km, while
strong lines are formed at heights from 100 km to 330 km.

Fig. 5 shows the effective heights of formation of Fe I lines in
the infrared and visible spectrum. It should be noted that the
weak iron lines ($d< 0.1$, $\lambda\lambda$ 1600--1800 nm, $EP=
6.4$~eV) are formed in the deepest photospheric layers. The
blueshifts of these deepest-formed infrared lines are not the
largest ones (see Fig 1a) due to the effect of additional
absorption from the lanes. Therefore the blueshifts of infrared
weakest lines are smaller than the blueshifts of the visible
weakest lines with $EP=4.25$~eV.

Thus, the infrared iron lines are mainly formed in the reversed
granulation zone. Therefore intergranular lanes play
considerable importance role in the formation of the cores of
infrared lines. This accounts for the observed decrease of
blueshifts of infrared spectrum as compared with the visible
spectrum.
  \begin{figure}
\centerline{
\includegraphics [scale=0.3]{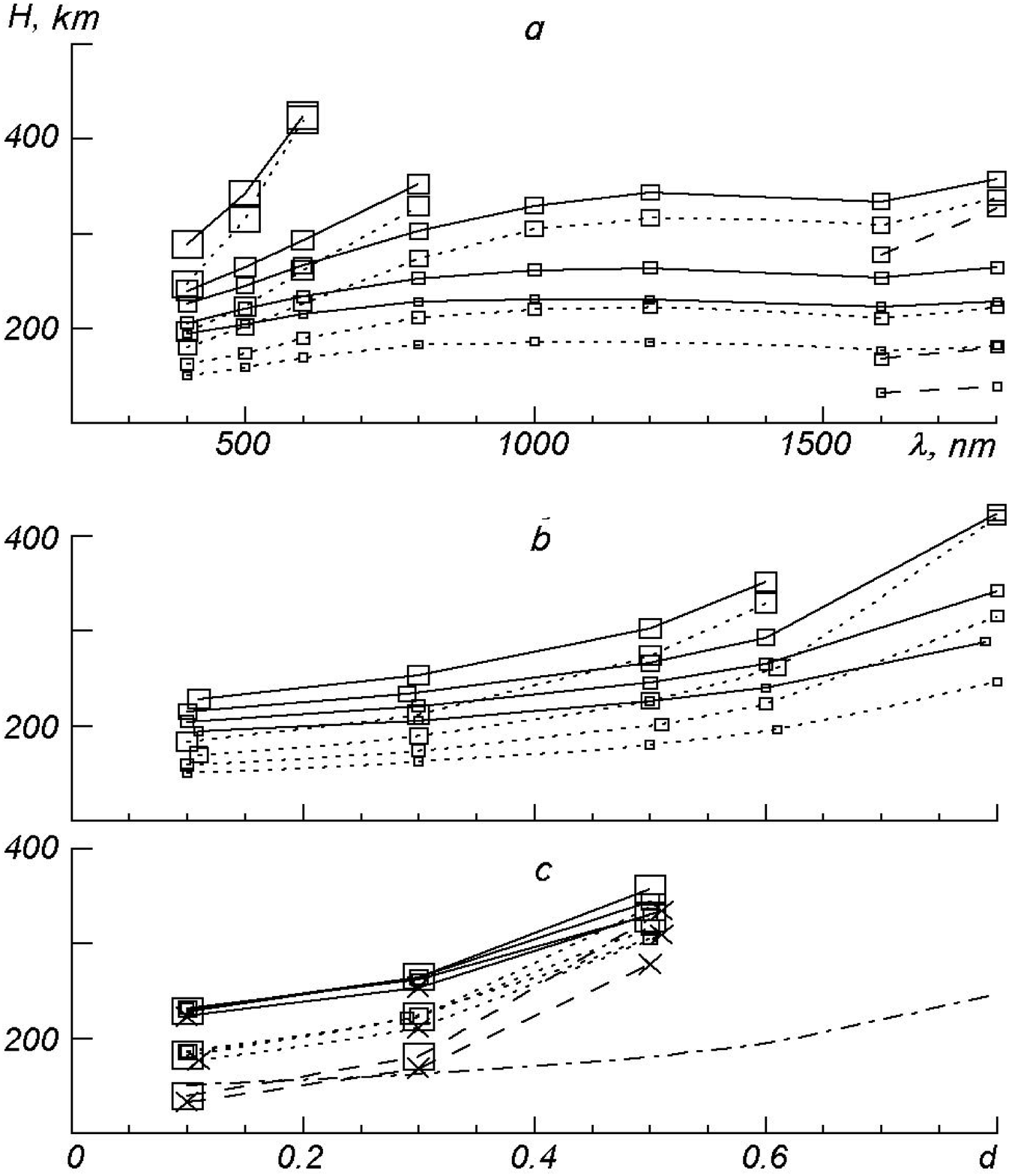}}
 \hfill
\vspace{0.1cm}
 \caption
{Effective heights of formation for artificial synthesized Fe I lines
depending on wavelength (a) and central depth for the visible
(b) and infrared (c) spectrum. Solid line: $EP=2.5$~eV, dotted
line: $EP=4.25$~eV, dashed line: $EP=6.4$~eV. Size of the
squares grows with line strength $d=0.1$, 0.3, 0.5, 0.6,
0.8 (a), and wavelength $\lambda=300$, 400, 500, 600, 800~nm (b)
and $\lambda=1000$, 1200, 1800~nm (c). Crosses mark
$\lambda=1600$~nm. Dot-and-dash line marks  $\lambda=400$~nm,
$EP=4.25$~eV shown for comparison.} \label{F-5}
 \end{figure}

\section{Conclusion}

We synthesized the Fe I lines in the framework of the 2-D HD
models of solar granulation and determined the wavelength shifts
of these lines and their dependence on the  line parameters. The
derived relations fit quite well the observation data.
Additionally we calculated the depression contribution functions
to demonstrate that the line shifts are determined by the
location of effective line formation region in convection cells.
Obtained  results  confirmed the convective character of the
line shifts.

Analysis of the line profiles and the corresponding contribution
functions showed that the contribution from granules and
intergranular lanes to the depression of spatially unresolved
spectral lines is essentially different for weak and strong
lines as well as for lines in the visible and infrared ranges.
In the visible range weak and moderate lines originate deep in
the photosphere, where the classical convection structure is not
disrupted, i.e., hot masses move upwards and cool masses
descend. Cores of strong iron lines are formed in the upper
photosphere, in the convection reversal zone, where  occurs a
temperature inversion as well as granulation inversion. The
temperature in the ascending central regions of a convection
cell is lower than in descending regions. The temperature
inversion increases the role of intergranular lanes in the
formation of the depression of the spatially averaged line. As a
result, the depression contribution from the regions with
downflows grows, and the line blueshifts diminish to zero or
even become red. In the infrared spectrum the effect of reserved
granulation on the line formation is stronger.  Due to growth of
line opacity with wavelength the level of effective line formation is
shifted higher into the photosphere  where are manifested the
reserved granulation zone. Therefore the line blueshifts in the
infrared spectrum are smaller than in the visible spectrum.

Convective motions cause the asymmetry  of spectral line
profiles. The blue wings and cores of spectral lines are mainly
formed in the central parts of granules, where upflows prevail.
The  red wings are mainly formed in intergranular lanes with
downflows. Greater  contributions from  intergranular lanes in
the depression of  strong lines results in the strengthening of
red wings of absorption lines and consequently a greater
asymmetry, as compared to weak and moderate lines. Therefore the
red wing of a spectral line can be regarded as an indicator of
the additional absorption  in intergranular lanes.

We have found that the weak infrared Fe~I lines with $\lambda
\approx$~1600 nm and very high excitation potentials (about 6.4
eV) are formed at the deepest photospheric layers. Such lines
may be of prime importance in diagnostics of stellar
atmospheres.




\begin{thebibliography}{99}

\bibitem{9}
C. Allende Prieto, R. J. Garcia Lopez, ``Fe I line shifts in
the optical spectrum of the Sun,'' Astron. and Astrophys. Suppl.
Ser., vol. 129, no. 1, pp. 41--44, 1998.


\bibitem{10}
I. N. Atroshchenko, A. S. Gadun, ``Three-dimensional
hydrodynamic models of solar granulation and their application
to a spectral analysis problem,'' Astron. and Astrophys., vol.
291, no. 2, pp. 635--656, 1994.

\bibitem{1}
P. N. Brandt, A. S. Gadun,  V. A. Sheminova, ``Absolute
shifts of Fe I and Fe II lines in solar active regions (disk
center),'' Kinematika i Fisika Nebes. Tel [Kinematics and
Physics of Celestial Bodies], vol. 13, no. 5, pp. 75--86, 1997.


\bibitem{11}
F. Cavallini, G. Ceppatelli, A. Righini, ``Profile variations of
some photospheric lines as observed in active regions across the
solar disk,''  Astron. and Astrophys.,
 vol. 205, no. 1/2, pp. 278--288, 1988.

\bibitem{12}
D. Dravins, B. Larsson,  \AA. Nordlund, ``Solar Fe II line
asymmetries and wavelength shifts,''  Astron. and Astrophys.,
vol. 158, no. 1/2, pp. 83--88, 1986.



\bibitem{13}
D. Dravins, L. N. Lindegren,  \AA. Nordlund, ``Solar
granulation: influence of convection on spectral line
asymmetries and wavelength shifts,''  Astron. and Astrophys.,
vol. 96, no. 1/2, pp. 345--364, 1981.

\bibitem{14}
B. Freytag, H. G. Ludwig,  M. Steffen, ``Hydrodynamical
models of stellar convection. The role of overshoot in DA white
dwarfs, A type stars, and the Sun,''  Astron. and Astrophys.,
vol. 313, no. 2, pp. 497--517, 1996.


\bibitem{2}
A. S. Gadun, ``Multidimensional hydrodynamic models of the solar
atmosphere: effects of radiative transfer in a multidimensional
perturbed medium,''  Kinematika i Fizika Nebes. Tel
[Kinematics and Physics of Celestial Bodies],
vol. 11, no. 3, pp. 54--72, 1995.


\bibitem{3}
A. S. Gadun, ``Iron abundance derived from two-dimensional
inhomogeneous solar model atmosphere. Fe I and Fe II lines
(center of the disk,''  Kinematika i Fizika Nebes. Tel
[Kinematics and Physics of Celestial Bodies],
vol. 12, no. 4, pp. 19--31, 1996.


\bibitem{4}
A. S. Gadun, ``Spatial variations in the Li I $\lambda$~671~nm
resonance line in two-dimensional simulated granulation,''
Kinematika i Fizika Nebes. Tel [Kinematics and Physics of
Celestial Bodies], vol. 15, no. 2, pp. 153--159, 1999.


\bibitem{6}
A. S. Gadun, A. Hanslmeier, ``Variations of line parameters
and bisectors over granular-intergranular regions in the 2-D
artificial solar granulation,'' Kinematika i Fizika Nebes. Tel
[Kinematics and Physics of Celestial Bodies], vol. 13, no. 3,
pp. 24--48, 1997.


\bibitem{7}
A. S. Gadun, A. Hanslmeier, ``Correlation analysis of
two-dimensional solar atmosphere,''  Kinematika i Fizika Nebes. Tel
[Kinematics and Physics of Celestial Bodies], vol. 16, no. 2, pp.
121--129, 2000.



\bibitem{8}
A. S. Gadun, A. Hanslmeier, ``Fe II lines in the problem of the
diagnostic of solar photospheric shocks,'' Kinematika i Fizika
Nebes. Tel [Kinematics and Physics of Celestial Bodies], vol.
16, no.2, pp. 130--137, 2000.

\bibitem{16}
A. S. Gadun, Ya. V. Pavlenko, ``1-D and 2-D model atmospheres
iron and lithium LTE abundances in the Sun,''  Astron. and
Astrophys., vol. 324, no. 1, pp. 281--288, 1997.

\bibitem{spansat}
A. S. Gadun,  V. A. Sheminova. SPANSAT: Program for Calculating
the LTE Absorption Line Profiles in Stellar Atmospheres, Kyiv,
1988, Inst. Theor. Phys., Academy of Sciences of UkrSSR,
Preprint No. ITF-88-87P.

\bibitem{5}
A. S. Gadun, Yu. Yu. Vorob'ev, ``Parameters of artificial
granules in two-dimensional hydrodynamic numerical simulation of
solar granulation,'' Astron. Zhurn., vol. 73, no. 4, pp.
623--632, 1996.


\bibitem{15}
A. S. Gadun, A. Hanslmeier,  K. N. Pikalov, ``Bisectors and
line-parameter variations over granular and intergranular
regions in 20D artificial granulation,''  Astron. and Astrophys.,
vol. 320, no. 3, pp. 1001--1012, 1999.

\bibitem{17}
A. S. Gadun, S. K. Solanki,  A. Johannesson, ``Granulation near
the solar limb: observations and 2-D modeling,'' in: Motions in
the Solar Atmosphere, pp. 201--204, Kluwer, Dordrecht, 1999.


\bibitem{18}
A. S. Gadun, S. K. Solanki, S. R. O. Ploner, et al.,
Scale-Dependent Properties of 2-D Artificial Solar Granulation,
Kiev, 1998. National Academy of Sciences of Ukraine, Main
Astronomical Observatory Preprint (MAO-98-4E).


\bibitem{19}
U. Grossman-Doerth, ``Height formation of solar photospheric
spectral lines,'' Astron. and Astrophys., vol. 285, no. 3, pp.
1012--1018, 1994.



\bibitem{20}
E. A. Gurtovrnko, V. A. Sheminova,  A. P. Sarychev, ``What is
the difference between ``emission`` and ``depression``
contribution functions?'' Solar Phys., vol. 136, no. 2, pp.
239--250, 1991.



\bibitem{21}
D. Kiselman, ``High-spectral resolution solar observations of
spectral lines used for abundance analysis,'' Astron. and
Astrophys. Suppl. Ser., vol. 104, no. 1, pp. 23--77, 1994.



\bibitem{22}
D. Kiselman, \AA. Nordlund, ``3D non-LTE line formation in the
solar photosphere and the solar oxygen line abundance,'' Astron.
and Astrophys., vol. 302, no. 2, pp. 578--586, 1995.



\bibitem{23}
A. Kucera, H. Balthasar, J. Rybak,  H. Wohl, ``Heights of
formation of Fe I photospheric lines,''  Astron. and Astrophys.,
vol. 332, no. 3, pp. 1069--1074, 1998.



\bibitem{24} R. L. Kurucz, Opacities for Stellar Atmospheres (CD ROM 2) 1993.


\bibitem{25}
P. Magain, ``Contribution function and the depths of formation
of spectral lines,''  Astron. and Astrophys., vol. 163, no. 1/2,
pp. 135--139, 1986.


\bibitem{26}
D. Nadeau, J. P. Mailard, ``Observational evidence of line
shifts induced by the convective overshoot in the atmospheres of
red giants,'' Astrophys. J., vol. 327, no. 1, pp. 321--327,
1988.


\bibitem{27}
\AA. Nordlund, D. Dravins, ``Solar granulation. III.
Hydrodynamic model atmospheres,'' Astron. and Astrophys., vol.
228, no. 1, pp. 155--183, 1990.


\bibitem{28}
K. Pushmann, A. Hanslmeier,  S. Solanki, Solar and Stellar
Granulations: IAU Symp. 178, K. G. Strassmeier (Editor), p. 117,
Vienna, 1995.


\bibitem{29}
J. Ramsauer, S. K. Solanki,  E. Biemont, ``Interesting lines
in infrared solar spectrum. II. Unblended lines between
$\lambda$ 1.0 and $\lambda$ 1.8 $\mu$m,'' Astron. and Astrophys.
Suppl. Ser., vol. 113, no. 1, pp. 71--89, 1995.


\bibitem{30}
M. P. Rast, \AA. Nordlund, R. F. Stein, J. Toomre,
``Ionization effects in three-dimensional solar granulation
simulations,'' Astrophys. J., vol. 408, no. 1, pp. L53--L56.


\bibitem{31}
I. Ruedi, S. K. Solanki, W. Livingston, J. Harvey,
``Interesting lines in infrared solar spectrum. III. A
polarimetric survey between 1.5 and 2.50 $\mu$m,'' Astron and
Astrophys. Suppl. Ser., vol. 113, no. 1, pp. 91--106, 1995.


\bibitem{32}
D. Samain, ``In the ultraviolet spectrum of the quiet Sun
redshifted?'', Astron. and Astrophys., vol. 244, no. 1, pp.
217--227, 1991.


\bibitem{33}
R. F. Stein, \AA. Nordlund, ``Simulations of solar granulation.
I. General properties,'' Astrophys. J., vol. 499, no. 2, pp.
914--933, 1998.


\end{thebibliography}
\end{document}